\RequirePackage{lineno}
\documentclass[prx,twocolumn,preprintnumbers,superscriptaddress]{revtex4} 

\usepackage{graphicx}
\usepackage{bm}
\usepackage{amsmath}
\usepackage{amssymb}
\usepackage{amsfonts}
\usepackage{fancyhdr}
\usepackage{slashed}
\usepackage{latexsym,epsfig,bbm}

\newcommand{\bra}[1]{\langle{#1}|}
\newcommand{\ket}[1]{|{#1}\rangle}

\newcommand{\braket}[2]{\langle{#1}|{#2}\rangle}

\usepackage{color}
\definecolor{blue}{rgb}{0,0.2,1}

\definecolor{red}{rgb}{0.9,0,0}

\newcommand{\past}[1]{\overleftarrow{#1}}
\newcommand{\fut}[1]{\overrightarrow{#1}}
\begin{document}

\title{Superior memory efficiency of quantum devices for the simulation of continuous-time stochastic processes}

\author{Thomas J. Elliott}
\email{physics@tjelliott.net}
\affiliation{School of Physical and Mathematical Sciences, Nanyang Technological University, Singapore 637371}
\affiliation{Complexity Institute, Nanyang Technological University, Singapore 637723}
\author{Mile Gu}
\email{mgu@quantumcomplexity.org}
\affiliation{School of Physical and Mathematical Sciences, Nanyang Technological University, Singapore 637371}
\affiliation{Complexity Institute, Nanyang Technological University, Singapore 637723}
\affiliation{Centre for Quantum Technologies, National University of Singapore, 3 Science Drive 2, Singapore 117543}

\date{\today}

\begin{abstract}
Continuous-time stochastic processes pervade everyday experience, and the simulation of models of these processes is of great utility. Classical models of systems operating in continuous-time must typically track an unbounded amount of information about past behaviour, even for relatively simple models, enforcing limits on precision due to the finite memory of the machine. However, quantum machines can require less information about the past than even their optimal classical counterparts to simulate the future of discrete-time processes, and we demonstrate that this advantage extends to the continuous-time regime. Moreover, we show that this reduction in the memory requirement can be unboundedly large, allowing for arbitrary precision even with a finite quantum memory. We provide a systematic method for finding superior quantum constructions, and a protocol for analogue simulation of continuous-time renewal processes with a quantum machine.
\end{abstract}
\maketitle 


Our experience of the world manifests as a series of observations. The goal of science is to provide a consistent explanation for these, and further, make predictions about future observations. That is, science aims to provide a \emph{model} of Nature, to describe the processes that give rise to the observations. It is possible to devise many different models that make identical predictions, and so it is desirable to have criteria that discern the `best' model. One such guiding philosophy is Occam's razor ``plurality should not be posited without necessity", which can be interpreted as requiring that a model should be the `simplest' that accurately describes our observations. 

This now leaves us with the question of how to determine the simplest model. The field of \emph{computational mechanics} \cite{crutchfield1989inferring, shalizi2001computational, crutchfield2012between} seeks to answer this, defining the optimal predictive model of a process to be that which requires the least information about the past in order to predict the future, and uses this minimal memory requirement as a measure of complexity. There is motivation for preferring simpler models beyond the inclination for elegance; it allows one to make more fundamental statements about the processes themselves, due to their irreducible nature \cite{crutchfield2012between}. More pragmatically, it also facilitates the building of \emph{simulators} (devices emulating the behaviour of the system [Fig.~1]) \cite{feynman1982simulating} for the process, as simpler models require fewer resources (here, internal memory).

\begin{figure}
\includegraphics[width=\linewidth]{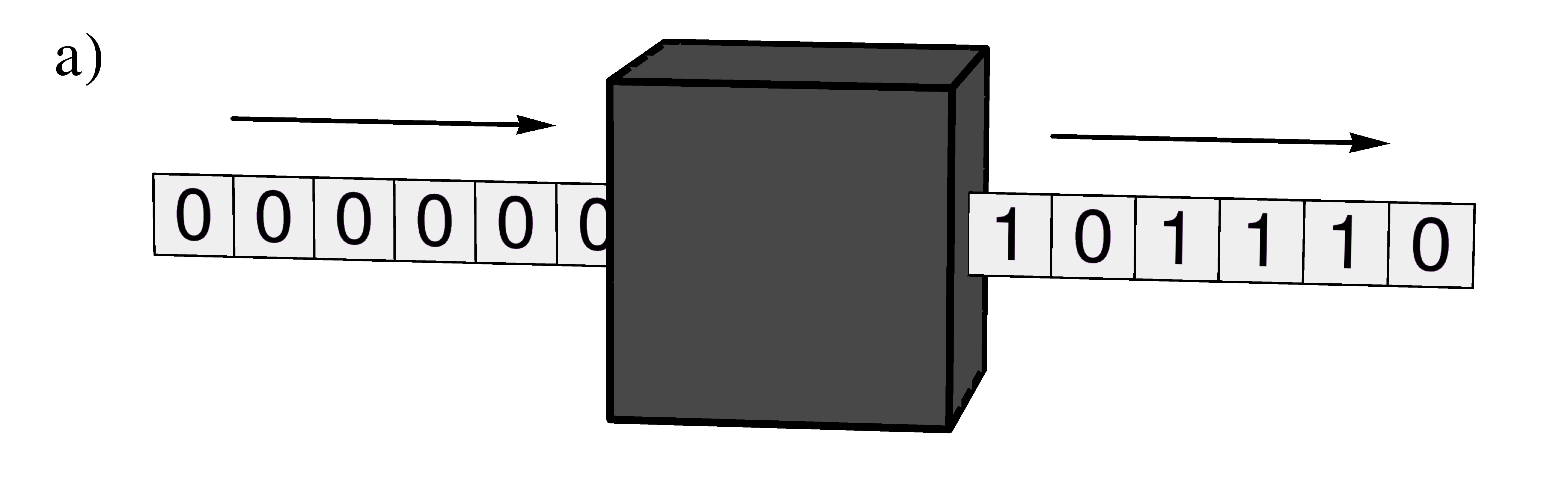}
\includegraphics[width=\linewidth]{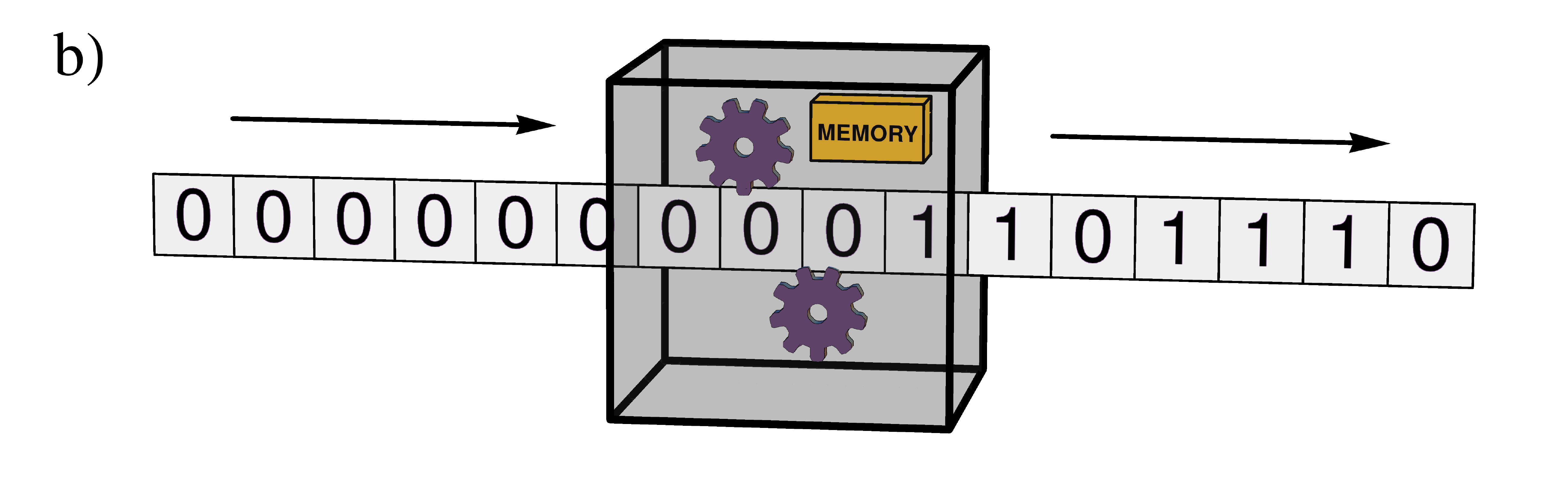}
\caption{{\bf Models and simulators.} (a) A system can be viewed as a black box which outputs what we observe, and our explanation of the observations forms a model. (b) We can build simulators to implement our models, in order to test them and make predictions about the future. Computational mechanics defines the optimal models to be those with the minimal internal memory requirement.}
\label{figsim}
\end{figure}

Discrete-time processes have been well-studied within the computational mechanics framework \cite{crutchfield1997statistical, tino1999extracting, palmer2000complexity, clarke2003application, park2007complexity, li2008multiscale, crutchfield2009time, lohr2009properties, haslinger2010computational, kelly2012new, garner2017thermodynamics, marzen2015informational}. However, it has recently been shown that quantum machines can be constructed that in general exhibit a lower complexity, and hence a lower memory requirement, than their optimal classical counterparts \cite{gu2012quantum, suen2015classical, mahoney2016occam, riechers2016minimized, aghamohammadi2016ambiguity, aghamohammadi2016extreme, garner2016unbounded, thompson2017using}. This substantiates the perhaps surprising notion that a quantum device can be more efficient than a classical system, even for the simulation of a purely classical stochastic process. This has recently been verified experimentally \cite{palsson2017experimentally}.

Continuous-time processes have also recently been the focus of computational mechanics studies \cite{riechers2016beyond, marzen2017informational}. While the principles of optimality for discrete-time processes can be directly exported to the continuous-time case, systematic study of the underlying architecture has only been carried out for a restricted set of processes; renewal processes \cite{marzen2017informational}. Renewal theory describes a generalisation of Poisson processes \cite{smith1958renewal, barbu2009semi}, where a system emits at a time drawn from a probabilistic distribution, before returning to its initial state (one can also view this as a series of events separated by probabilistic dwell times). Despite their apparent simplicity, such processes have many applications, including models of lifetimes \cite{doob1948renewal}, queues \cite{kalashnikov2013mathematical}, and neural spike trains \cite{gerstner2002spiking, haslinger2010computational, crutchfield2015time}.

Here, we show that the quantum advantage can be extended to continuous-time processes. Focussing on renewal processes, we provide a systematic construction for determining quantum machines that require less information about the past for accurate future prediction than the optimal classical models. We provide a protocol that can be used to implement such quantum machines as analogue simulators of a renewal process. We then illustrate the quantum advantage with two examples. In particular, we find that while the classical machines typically need infinite memory, a quantum machine can in some cases require only finite memory. We conclude by arguing that the finite memory requirement may be a typical property of quantum machines, suggest possible mediums for a physical realisation of the quantum machines, and discuss prospects for future work. A technical appendix is provided with additional details on the framework used, and the derivations for the examples.


{\bf Framework.} Here we review the computational mechanics framework used to express our results (a more complete overview may be found elsewhere \cite{shalizi2001computational, crutchfield2012between}). We consider continuous-time discrete-alphabet stochastic point processes. Such a process $\mathcal{P}$ is characterised by a sequence of observations $(x_{\bm{n}},t_{\bm{n}})$, drawn from a probability distribution $P(X_{\bm{n}},T_{\bm{n}})$ \cite{khintchine1934korrelationstheorie}. Here, the $x_n$, drawn from an alphabet $\mathcal{A}_n$, are the symbols emitted by the process, while the $t_n$ record the times between emissions $n-1$ and $n$. For shorthand, we denote the dual ${\bm x}_n=(x_n,t_n)$, and similarly $\bm{X}_{n}$ for the associated stochastic variable. We denote a contiguous string of observations of emitted symbols and their temporal separations by the concatenation ${\bm x}_{l:m}={\bm x}_l{\bm x}_{l+1}\ldots{\bm x}_{m-1}$, and for a stationary process we mandate that $P({\bm X}_{0:L})=P({\bm X}_{s:s+L})\forall s,L\in\mathbb{Z}$. Note that the discrete-time case consists of either coarse-graining the $t_n$, or considering processes where such dwell times are either identical or irrelevant.

We define the past of a process $\past{\bm{x}}={\bm x}_{-\infty:0}(\emptyset,t_{0^+})$, where 0 is the current emission step (i.e.~the next emitted symbol will be $x_0$, and $\emptyset$ denotes that this symbol is currently undetermined), and $t_{0^+}$ is the time since the last emission, with associated random variable $T_{0^+}$. Analogously, defining $t_{0^-}$ as the time to the next emission, we can denote the future $\fut{\bm{x}}=(x_0,t_{0^-}){\bm x}_{1:\infty}$ \cite{marzen2017informational}. The causal states of the process are then an equivalence class defined according to a predictive equivalence relation \cite{crutchfield1989inferring}; two past sequences $\past{{\bm x}}$ and $\past{{\bm x}}'$ belong to the same causal state (i.e. $\past{{\bm x}}\sim_e\past{{\bm x}}'$) iff they satisfy
\begin{equation}
P(\fut{{\bm X}}|\past{{\bm X}}=\past{{\bm x}})=P(\fut{{\bm X}}|\past{{\bm X}}=\past{{\bm x}}').
\end{equation}
We use the notation $S_j$ to represent the causal state labelled by some index $j$.

We desire models that are predictive, wherein the internal memory of a simulator implementing the model contains all (and no additional) information relevant to the future statistics that can be obtained from the entire past. The first part of this entails the simulator memory having the same predictive power as knowledge of the entire past (prescience \cite{shalizi2001computational}), while the second ensures that knowledge of the memory provides no further predictive power than observing the entire past output (information about the future accessible in this manner is referred to as oracular \cite{crutchfield2010synchronization}, and implies the simulator having decided aspects of its future output in advance). This notion of predictive models is stricter than the broader class of generative models, which must only be able to faithfully reproduce future statistics; internal states of models in the broader class may contain additional information that allows for better prediction of future outputs than knowledge of the past, violating the non-oracular condition. We note that while there exist generative models that can operate with lower memory than the optimal predictive models we will now introduce, as this is achieved by leveraging oracular information we do not consider such models here.

 The provably optimal predictive classical models, termed `$\varepsilon$-machines', operate on the causal states \cite{crutchfield1989inferring, shalizi2001computational}.  In general the systematic structure of these models is well-understood only for discrete-time processes, though as we later discuss recent efforts have been made towards constructing corresponding continuous-time machines. A discrete-time $\varepsilon$-machine may be represented by an edge-emitting hidden Markov model, in which the hidden states are the causal states, the transitions (edges) between these states involve the emission of a symbol from the process alphabet, and the string of emitted symbols forms the process. The edges are defined by a dynamic $T_{kj}^{(x)}$ describing the probability of transitioning from causal state $S_j$ to $S_k$ while emitting symbol $x$. The $T_{kj}^{(x)}$ are thus defined by the statistics of the process, and because they depend only on the current hidden state the model is Markovian. Further, as the predictive equivalence relation ensures that the system is always in a definite causal state defined wholly and uniquely by its past output, $\varepsilon$-machines are unifilar \cite{shalizi2001computational}. This means that for a given initial causal state and subsequent emission(s), the current causal state is known with certainty.

The quantity of interest for our study is the \emph{statistical complexity} $C_\mu$, which answers the question ``What is the minimal information required about the past in order to accurately predict the future?". It is defined as the Shannon entropy \cite{nielsen2000quantum} of the steady state distribution $\pi$ of the causal states $S_j$;
\begin{equation}
\label{eqcmudisc}
C_\mu=-\sum_j\pi(S_j)\log_2(\pi(S_j)).
\end{equation}
The use of Shannon entropy is motivated by considering the memory to be the average information stored about the past (alternatively, it can be viewed as the average information communicated in the process from the past to the future). Due to the ergodic nature of the processes considered, the time average and the ensemble average are equivalent. However, one could also consider the Hartley entropy, that is, the size of the substrate into which the memory is encoded (i.e.~the logarithm of the number of states) \cite{crutchfield1989inferring}. It can be shown that the $\varepsilon$-machine also optimises this measure \cite{shalizi2001computational}, though we shall here focus on the former measure, and implicitly consider an ensemble scenario. That is, when operating $N$ independent simulators, the total memory required tends to $NC_\mu$ as $N\to \infty$ \cite{nielsen2000quantum}. The statistical complexity is lower-bounded by the mutual information between the past and future of the process, referred to as the excess entropy $E=I(\past{{\bm X}};\fut{{\bm X}})$ \cite{shalizi2001computational}. 

Although the predictive equivalence relation defines the optimal model for both discrete- and continuous-time processes, as noted earlier, most works so far have been devoted to studying the $\varepsilon$-machines of discrete-time processes. It is only recently that a similar systematic causal architecture has been uncovered for a restricted set of continuous-time processes, renewal processes \cite{marzen2017informational}. Renewal processes form a special case of the above, where each emission occurs at an independent and identically distributed (IID) probabilistic time, and emits the same symbol. Such processes are defined entirely by this emission probability density $\phi(t)$, and the sequence is fully described by the emission times alone. It is useful to define the following quantities for a renewal process: the survival probability $\Phi(t)=\int_t^\infty \phi(t')dt'$; and the mean firing rate $\mu=(\int_0^\infty t\phi(t)dt)^{-1}$.

\begin{figure}
\includegraphics[width=0.75\linewidth]{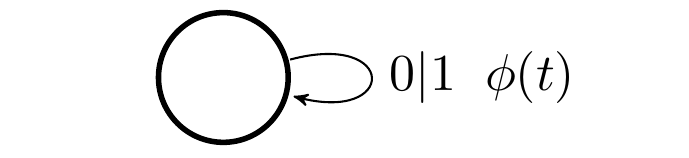}
\caption{{\bf Generative model for a renewal process.} Diagram depecting a generative model for a renewal process. The labelling indicates that a symbol 0 is emitted with probability 1, at time $t$ with probability density $\phi(t)$, and returns to the same state.}
\label{figrenewal}
\end{figure}

In Fig.~2 we show a generative model for such a process. Because of the IID nature of the process, the only relevant part of the past in predicting the future statistics is the time since the last emission $t_{0^+}$, and this assists us only in predicting the time to the next emission $t_{0^-}$ \cite{marzen2017informational}. Thus, the causal equivalence relation simplifies to
\begin{equation}
\label{eqcausalequiv}
t_{0^+}\sim_e t_{0^+}'\Leftrightarrow P(T_{0^-}|T_{0^+}=t_{0^+})=P(T_{0^-}|T_{0^+}=t_{0^+}').
\end{equation}
We label the causal states $S_{t_{0^+}}$ according to the minimum $t_{0^+}$ belonging to the equivalence class. Depending on the form of $\phi(t)$, we can determine which $t_{0^+}$ belong to the same causal state. Notably, if $\phi(t)$ is Poissonian, the time since the last emission is irrelevant (as the decay rate is constant), and hence all $t_{0^+}$ belong to the same causal state - the process is memoryless and has $C_\mu=0$. All other processes involve a continuum of causal states, which may either extend indefinitely, terminate in a single state at a certain time, or eventually enter a periodic continuum (see Appendix A). The steady state probability density $\pi(S_t)$ of the causal states depends on this causal architecture (Appendix B). We specifically highlight that states in the initial continuum have $\pi(S_t)=\mu\Phi(t)$; as we will later discuss, this is the only necessary part of the architecture once we turn to quantum causal states.

The statistical complexity of the process can be defined in correspondence with Eq.~\eqref{eqcmudisc}, by taking the continuous limit of a discretised analogue of the process; 
\begin{equation}
\label{eqcmucont}
C_\mu=\lim_{\delta t\to0}-\sum_{n=0}^\infty \pi(S_{n\delta t})\delta t\log_2(\pi(S_{n\delta t})\delta t).
\end{equation}
This quantity will however either be zero (for a Poissonian emission probability density), or infinite (for all other distributions), due to the infinitesimal coarse-graining. Classically therefore, it is not the most enlightening measure of complexity, and has motivated earlier work on this topic \cite{marzen2017informational} to instead consider use of the differential entropy for the statistical complexity; $C_\mu^{(\mathrm{DE})}=-\int_0^\infty dt\pi(S_t))\log_2\pi(S_t)$. While this quantity allows for a comparison of the complexity of two processes, we find it lacking as an absolute measure of complexity, as it requires one to take logarithms of dimensionful quantities, and loses the original physical motivation of being the information contained within the process about its past. Instead, we will employ the true continuum limit of the Shannon entropy Eq.~\eqref{eqcmucont} as the measure of a process' statistical complexity, accepting the infinities as faithfully representing that classical implementations of such models do indeed require infinite memory.


{\bf Quantum causal states.} It has been shown that a quantum device simulating a discrete-time process can in general require less memory than the optimal classical model \cite{gu2012quantum}. In order to assemble such a device, for each causal state $S_j$ one must construct a corresponding quantum causal state $\ket{S_j}=\sum_{xk}\sqrt{T_{kj}^{(x)}}\ket{x}\ket{k}$, where, as defined above, the transition dynamic $T_{kj}^{(x)}$ is the probability that a system in $S_j$ will transition to $S_k$, while emitting symbol $x$. The machine then operates by mapping the state $\ket{k}$ with a blank ancilla to $\ket{S_k}$, following which measurement of the $\ket{x}$ subspace will produce symbol $x$ with the correct probability, while leaving the remaining part of the system in $\ket{S_k}$. The internal steady state of the machine is given by $\rho=\sum_j \pi(S_j) \ket{S_j}\bra{S_j}$. We refer to such constructions as q-machines, and their internal memory $C_q$ can be described by the von Neumann entropy \cite{nielsen2000quantum} of the steady state;
\begin{equation}
C_q=-\mathrm{Tr}(\rho\log_2\rho).
\end{equation}
Unlike classical causal states, the overlaps $\braket{S_j}{S_k}$ of different quantum causal states are in general non-zero, and hence $C_q\leq C_\mu$ (typically the inequality is strict); thence, the q-machine has a lower internal memory requirement than the corresponding $\varepsilon$-machine \cite{gu2012quantum}. Physically, this memory saving can be understood as the lack of a need to store information that allows complete discrimination between two pasts when they have some overlap in their conditional futures. This entropy reduction acquires operational significance when one considers an ensemble of independent simulators of a process sharing a common total memory \cite{gu2012quantum}. As with the classical case, $C_q$ is also lower bounded by the excess entropy of the process. Note that while this quantum construction is superior to the optimal classical model, it does not necessarily provide the optimal quantum model. Indeed, for particular classes of process, constructions involving several symbol outputs are known that have even lower internal memory \cite{mahoney2016occam, riechers2016minimized}, and there may exist as yet unknown further optimisations beyond this. Such known improvements however are not relevant for the processes we consider. 

We now seek to extend this quantum memory reduction advantage to the realm of continuous-time processes. To do so, we first define a wavefunction $\psi(t)=\sqrt{\phi(t)}$. We can rephrase the survival probability and mean firing rate in terms of this wavefunction: $\Phi(t)=\int_t^\infty |\psi(t')|^2dt'$; and $\mu=(\int_0^\infty t|\psi(t)|^2dt)^{-1}$. Inspired by the quantum construction for discrete-time processes, we wish to construct quantum causal states $\ket{S_t}$ such that when a measurement is made of the state (in a predefined basis), it reports a value $t'$ with probability (density) $P(T_{0^-}=t'|T_{0^+}=t)$. We may view the quantum causal state as a continuous alphabet (representing the value of $t_{0^-}$) analogue of the discrete case, with only a single causal state ($S_0$) the system may transition to after emitting this symbol.

The probability density $P(T_{0^-}=t'|T_{0^+}=t)$ is given by $\phi(t+t')/\int_t^\infty \phi(t')dt'=\phi(t+t')/\Phi(t)$. By analogy with the discrete case we construct our quantum causal states as $\ket{S_t}=\int_0^\infty dt' \sqrt{P(T_{0^-}=t'|T_{0^+}=t)}\ket{t'}$, and thus:
\begin{equation}
\label{eqquantumcausal}
\ket{S_t}=\frac{1}{\sqrt{\Phi(t)}}\int_0^\infty dt'\psi(t+t')\ket{t'}.
\end{equation}
We emphasise that while the wavefunction is encoding information about time in the modelled process, the q-machine used for simulation may encode it in any practicable continuous variable, such as the position of a particle. The measurement basis used to obtain the correct statistics is of course that defined by $\{\ket{t}\}$ (that is, measurement outcome $t'$ occurs with probability density $|\braket{t'}{S_t}|^2=|\psi(t+t')|^2/\Phi(t)$ when the system is in state $\ket{S_t}$).

When the first segment $[0,\tilde{t})$ of the continuous variable in a quantum causal state is swept across, if the system is not found to be in this region the state is modified by application of the projector $\Pi_{\tilde{t}}=\int_{\tilde{t}}^\infty dt\ket{t}\bra{t}$ and appropriate renormalisation. When this projector is applied to the state $\ket{S_t}$, the resulting state is simply $\ket{S_{t+\tilde{t}}}$ displaced by $\tilde{t}$; by correcting for this displacement the effect of the measurement sweep is exactly identical to the change in the internal memory of the machine if no emission is observed in a time period $\tilde{t}$, and thus the quantum causal states automatically update when measurement sweeps are used to simulate the progression of time.

The overlap of two quantum causal states can straightforwardly be calculated:
\begin{equation}
\label{eqstateoverlap}
\braket{S_{a}}{S_{b}}=\frac{1}{\sqrt{\Phi(a)\Phi(b)}}\int_0^\infty dt\psi(t+a)\psi(t+b).
\end{equation}
By their very construction, these quantum states will automatically merge states with identical future statistics, even if we neglect the underlying causal architecture. Recall the causal equivalence relation Eq.~\eqref{eqcausalequiv}. Since these probabilities wholly define the quantum states, if two quantum states have the same future statistics they are identical by definition. Due to the linearity of quantum mechanics, the steady state probabilities of the identical quantum states are added together to find the total probability for the state, much the same way as the underlying state probabilities are added together when merging states to form the classical causal states. Thus, when constructing the quantum `causal' states, we are at liberty to ignore the classical causal architecture as described in Appendix A, without any penalty to the information that is stored by the q-machine, and instead construct quantum states for all $t\geq0$ according to the prescription of Eq.~\eqref{eqquantumcausal}. Note that the causal architecture can still be used as a calculational aid.


{\bf Memory of continuous-time q-machines.} From Eq.~\eqref{eqstateoverlap} we see that in general the overlaps of the quantum causal states are non-zero, unlike the corresponding classical states, which are orthogonal. Because of this reduced distinguishability of the quantum causal states, the entropy of their steady state distribution is less than that of the classical causal states, and hence the amount of information that must be stored by the q-machine to accurately predict future statistics is less than that of the optimal classical machine, evincing a quantum advantage for the simulation of continuous-time stochastic processes. We will later show with our examples that this advantage can be unbounded, wherein q-machines have only a finite memory requirement for the simulation of processes for which the $\varepsilon$-machine requires an infinite amount of information about the past. Note that even when we consider coarse-graining the time since the last emission to a resolution of finite intervals $\delta t$ we shall still see a quantum advantage due to the non-orthogonality of the quantum states. Note also that decoherence of the memory into the measurement basis destroys the quantum advantage, and will result in the classical internal memory cost $C_\mu$ (see Appendix C).

The density matrix describing the internal state of the q-machine is given by $\rho = \int_0^\infty dt \pi(S_t) \ket{S_t} \bra{S_t}.$ As discussed above, we can construct the quantum states $\ket{S_t}$ for all $t$, in which case their steady state probability density $\pi(S_t)$ is given by $\mu\Phi(t)$. We thus find that the elements of the density matrix are given by $\rho(a,b)=\mu\int_0^\infty dt\psi(t+a)\psi(t+b)$. From this, we can construct a characteristic equation to find the eigenvalues $\lambda_n$ that diagonalise the density matrix:
\begin{equation}
\label{eqcharacteristic}
\mu\int_0^\infty db\int_0^\infty dt\psi(t+a)\psi(t+b)f_n(b)=\lambda_nf_n(a).
\end{equation}
The information stored by the q-machine can then be expressed in terms of these eigenvalues; $C_q=-\sum_n\lambda_n\log_2\lambda_n$. We find that this quantity is invariant under rescaling of the time variable in the emission probability density (see Appendix D for details).


{\bf Building q-machine simulators of renewal processes.} While we have explained in the abstract sense how one constructs the quantum causal states, it is interesting to also consider the structure of a device that would actually perform such simulations. In fact, a digital simulation of the process, that simply emits a sequence $t_{0^-:L}$ on demand drawn from the correct probability distribution $P(T_{0^-:L}=t_{0^-:L}|T_{0^+}=t_{0^+})$ would be very straightforward to assemble in principle: one must prepare the state $\ket{S_{t_{0^+}}}$, and $L-1$ copies of $\ket{S_0}$ (the states are all independent due to the renewal process emissions being IID). Measurement of the first state provides the $t_{0^-}$, while measurement of the others provides the $t_{1:L}$. Because of the self-updating nature of the quantum causal states under partial measurement sweeps $[0,\tilde{t})$, measurement over such a range can be used to simulate the effect of waiting for a time $\tilde{t}$ for an emission.

However, this scheme is unsatisfactory as one must manually switch to a new state after each emission. Rather, a device that automatically begins operating on the state for the next emission after the previous state is finished would be preferable. We now describe such a construction, and even go a step further, by devising a setup that enables an analogue simulation of the process, and is thus able to provide emission times in (scaled) real time. For illustrative purposes, we first describe the protocol for discrete timesteps (that may be coarse-grained arbitrarily finely), and then discuss how it can be performed in continuous-time.

\begin{figure}
\includegraphics[width=\linewidth]{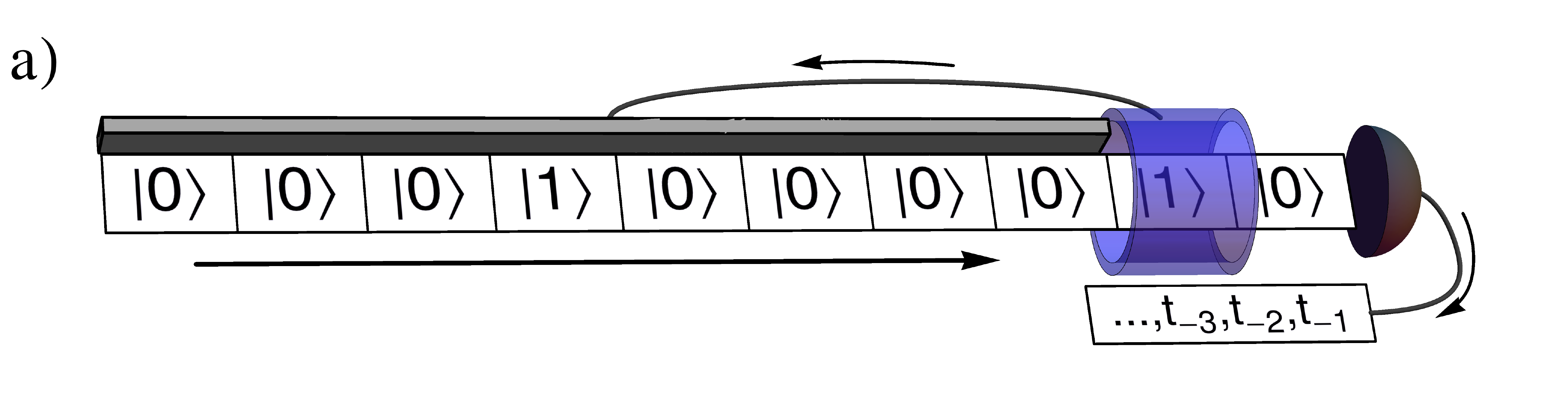}
\includegraphics[width=\linewidth]{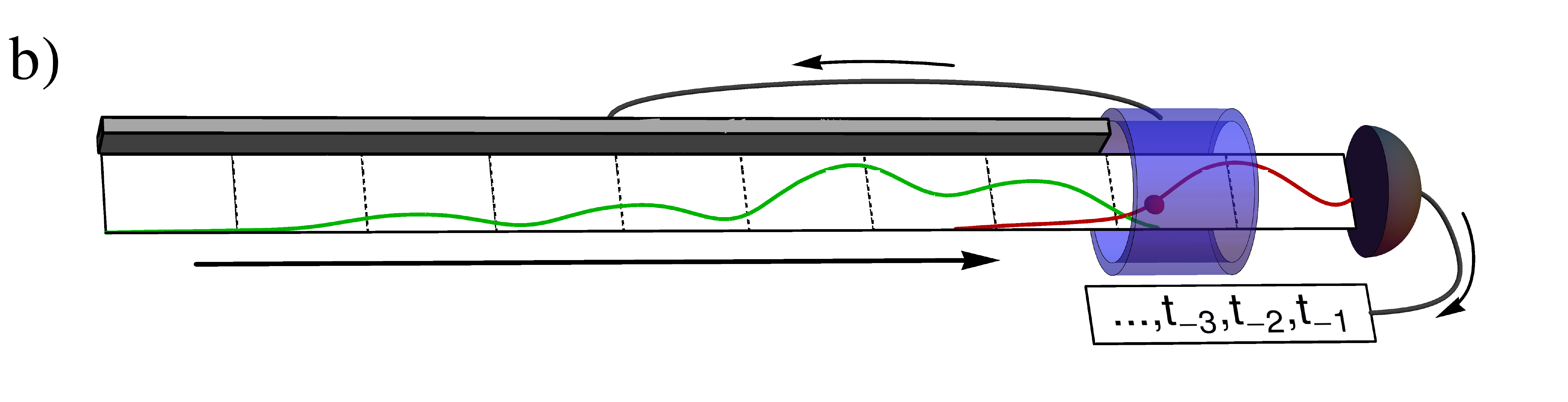}
\caption{{\bf q-machine simulators of renewal processes.} (a) Analogue simulator for a discrete-time renewal process, where a continuous chain of qubits is used to encode the quantum causal state. The simulator sweeps along the chain and alters the future of the chain conditional on the current qubit, with the mappings $\ket{0}\ket{1_n}\to\ket{0}\ket{1_n}$ and $\ket{1}\ket{0}^{\otimes\infty}\to\ket{1}\ket{\sigma_0}$. Measurement of the qubit state signifies whether an emission occurs in a given timestep. (b) Analogue simulator for continuous-time renewal processes, where the quantum causal state is encoded into the position of a particle. The simulator sweeps along this position and generates additional particles encoding future emissions conditional on the presence of the particle. Detection of the particle signals an emission event.}
\label{figcircuit}
\end{figure}

The procedure for the discrete-time case is as follows. Consider an infinite chain of qubits (two state quantum systems) labelled from 0 to $\infty$. Using $\ket{1_n}$ to denote the state where all qubits are in state $\ket{0}$ apart from the $n$th, which is in state $\ket{1}$, we can express the discretised analogues $\ket{\sigma_t}$ of the quantum causal states $\ket{S_t}$ as $\ket{\sigma_t}=\sum_n  \sqrt{P(T_{0^-}=n\delta t|T_{0^+}=t)\delta t}\ket{1_n}$, where $P(T_{0^-}=n\delta t|T_{0^+}=t)\to\phi(t+n\delta t)/\Phi(t)$ as $\delta t\to0$. The location $n$ of the qubit in state $\ket{1}$ then represents the time $n\delta t$ at which the emission occurs. We initialise the system in state $\ket{\sigma_{t_{0^+}}}$, according to the desired initial $t_{0^+}$. The chain is then processed sequentially, one qubit at a time, by performing a control gate on the qubit, which has the effect of mapping the next block of the chain to the state $\ket{\sigma_0}$ if the qubit is in state $\ket{1}$, and doing nothing otherwise (explicitly, the mapping required is $\ket{0}\ket{1_n}\to\ket{0}\ket{1_n}$ $\forall n\in\mathbb{Z}^+$ and $\ket{1}\ket{0}^{\otimes\infty}\to\ket{1}\ket{\sigma_0}$, where by construction these are the only possible input states). The qubit is then ejected from the machine (where measurement can be used to determine whether an emission event occurs at this time), and the machine then acts on the next qubit in the chain [Fig.~3(a)]. This operation has the effect of preparing the chain in a state that provides the correct conditional probabilities if no emission is observed, and prepares the state with the correct distribution for the next emission step if an emission is observed.

To operate this protocol in continuous-time, instead of encoding the state onto a discrete chain, we instead use a continuous degree of freedom, such as spatial position (henceforth referred to as the `tape'). As with the discrete case, we process sequentially along the tape, performing a unitary gate on the future of the tape, controlled on the current segment. Each emission step has its emission time encoded by the position of a particle on the tape [Fig.~3(b)]; the first particle on the tape is initialised in $\ket{S_t}=(1/\sqrt{\Phi(t)})\int_0^\infty dx\psi(t+x)\ket{x}$, where $x$ labels the position on the tape. Since the controlled unitary operation must be performed in discrete time, on a discrete length of tape, it is designed such that it acts, controlled on the presence of a particle in the block, by placing a particle in state $\ket{S_0}$, displaced to have its zero at the location of the control particle, and does nothing otherwise, akin to the discrete case above (that is, if the present particle is at position $x$, the combined state of the old and new particle is mapped to $\ket{x}\ket{S_{-x}}$, where we clarify that $\psi(t)=0$ if $t$ is negative). More formally, this can be written as the transformation $\int_0^\infty dt \int_L dx \psi(x+t)a^\dagger_{x+t}a^\dagger_xa_x$, where $L$ is the block of tape upon which the gate acts, and $a^\dagger_x$ creates a particle at $x$. Strictly, the gate should act in a nested fashion, by further generating an additional particle in an appropriately displaced state, when the new particle is placed within the current block. The machine then progresses to perform the same operation contiguously on the next block, while feeding out the previous block (equivalently, the tape can be fed through a static machine). Measurement of the positions of particles on the tape fed out then provides the simulated emission times.


{\bf Examples.} We illustrate our proposal with two examples. We show for both these examples that not only is there a reduction in the memory requirement of the q-machine compared to the $\varepsilon$-machine, but also that the q-machine needs only a finite amount of memory, while the classical has infinite memory usage. Here we summarise the results, and the technical details may be found in Appendices E and F.

\begin{figure}
\includegraphics[width=\linewidth]{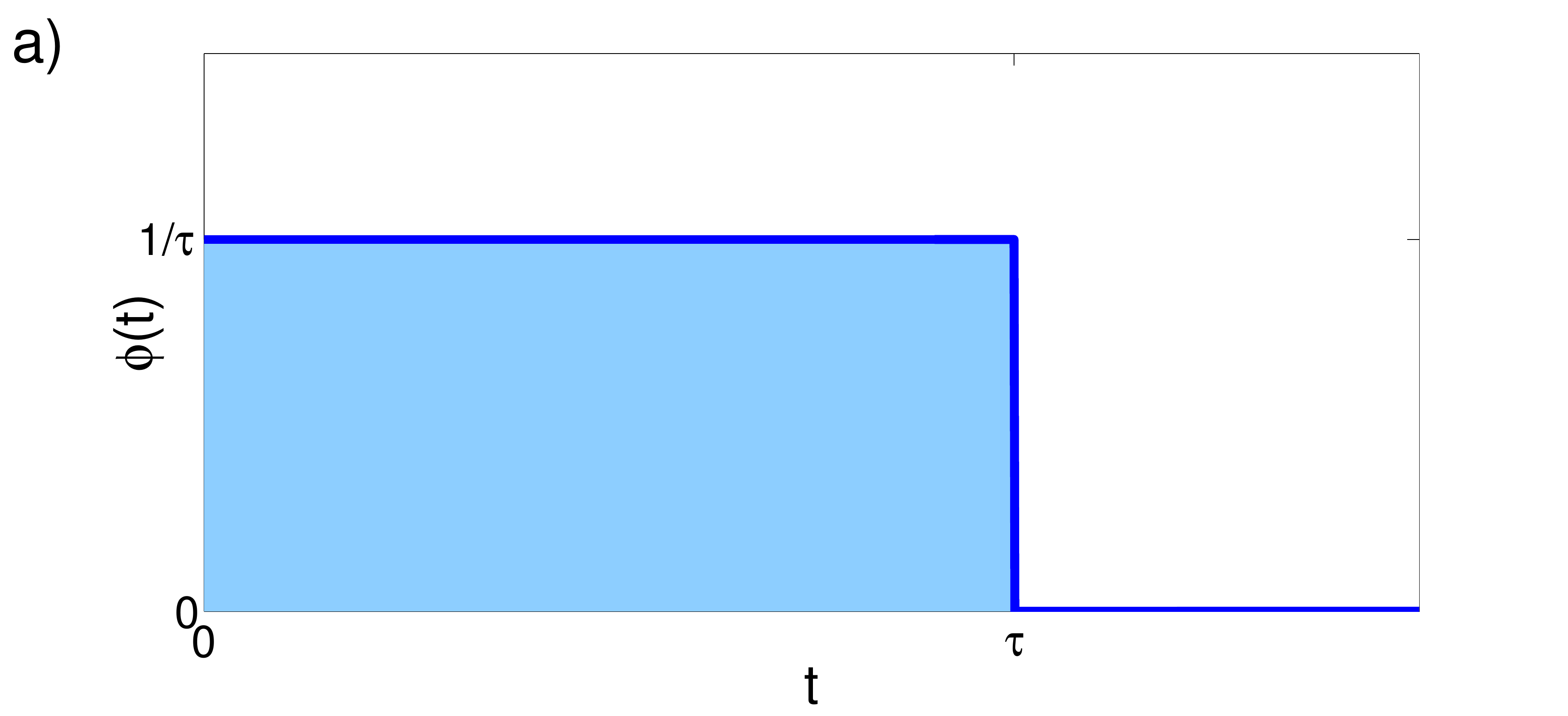}
\includegraphics[width=\linewidth]{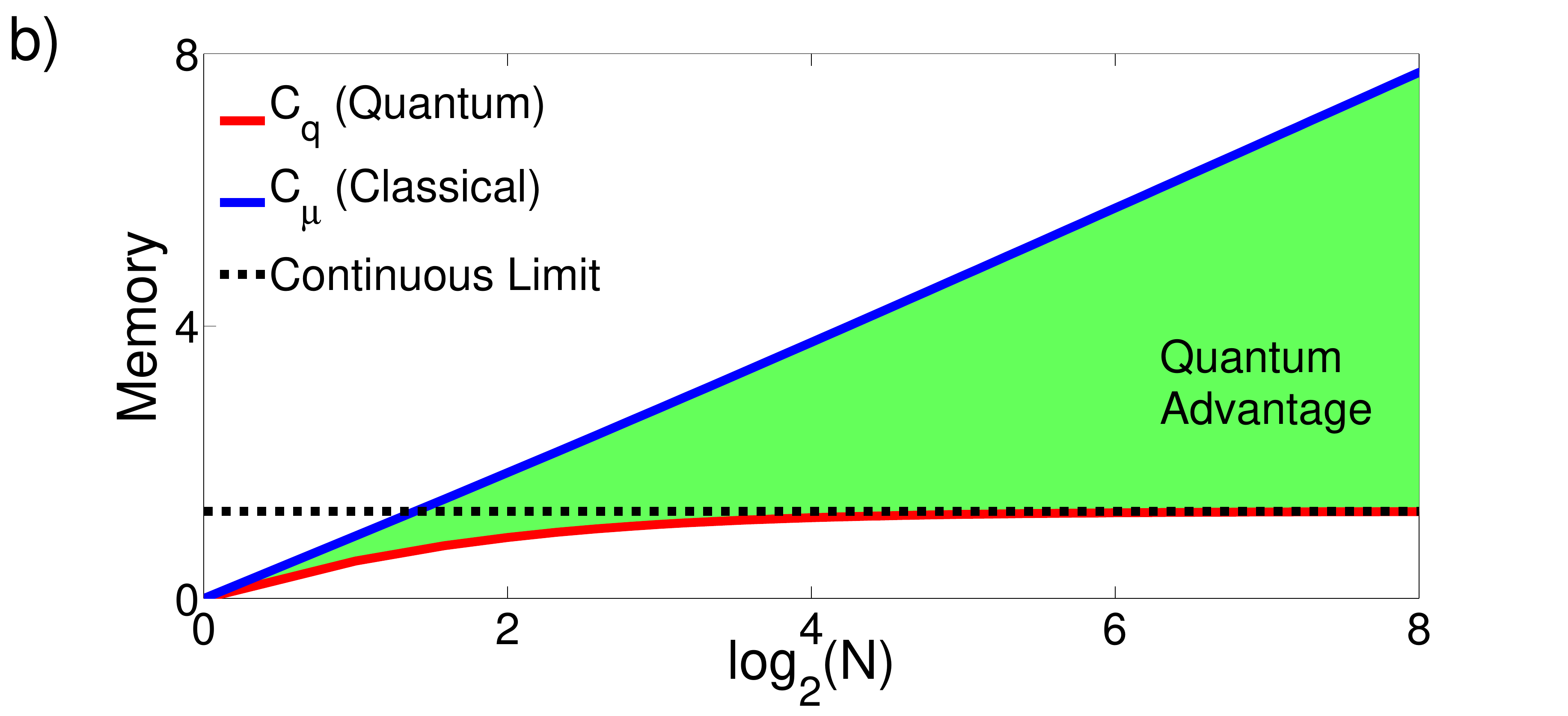}
\caption{{\bf Uniform emission probability.} (a) The corresponding emission probability density for a process with uniform emission probability in an interval $[0,\tau)$. (b) The classical memory $C_\mu$ required to simulate the process diverges logarithmically as the discretisation becomes finer ($N$ states), while the quantum memory $C_q$ converges on a finite value.}
\label{figuniform}
\end{figure}

The first example is a uniform emission probability over the interval $[0,\tau)$. The corresponding emission probability density is $\phi(t)=1/\tau$ for $0\leq t<\tau$, and zero elsewhere [Fig.~4(a)]. The corresponding mean firing rate and survival probability are given by $\mu=2/\tau$ and $\Phi(t)=1-t/\tau\;(t<\tau)$ respectively. The corresponding quantum causal states are given by $\ket{S_t}=\int_0^{\tau-t}dt'(1/\sqrt{\tau-t})\ket{t'}$, and we can solve Eq.~\eqref{eqcharacteristic} to find that $\lambda_n=8/(\pi(2n-1))^2$ for $n\in\mathbb{Z}^+$. We can use an integral test (see Appendix E) to show that $C_q=-\sum_{n=1}^{\infty}\lambda_n\log_2\lambda_n$ is bounded, and moreover, that $C_q\approx 1.2809$. In Fig.~4(b) we show how the memory required by the q-machine tends towards this value as we use an increasingly fine coarse-graining of the discretised analogue of the process to approach the continuous limit, while the memory needed by the optimal classical machine diverges logarithmically. The memory requirement exceeds the lower bound set by the excess entropy $E=\log_2e-1\approx0.4427$.

\begin{figure}
\includegraphics[width=\linewidth]{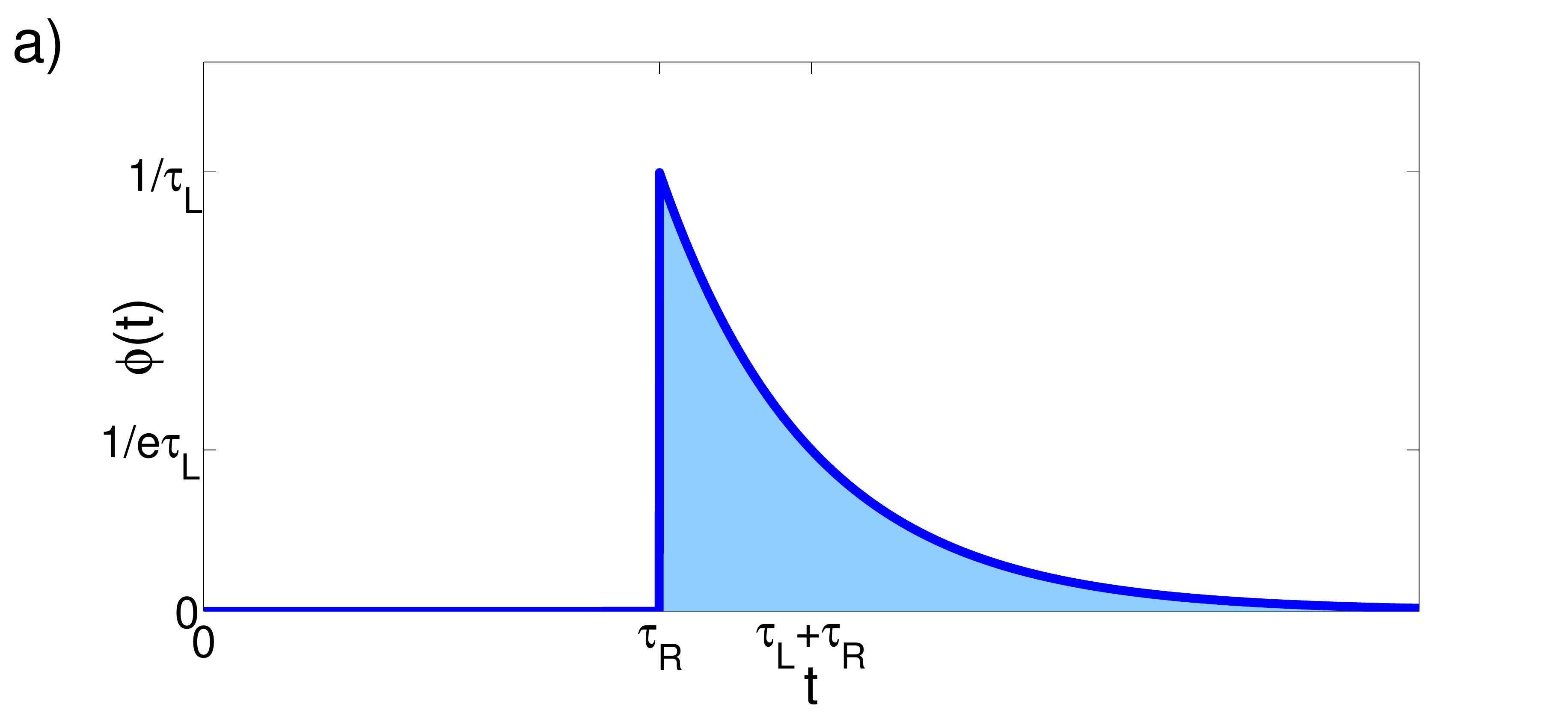}
\includegraphics[width=\linewidth]{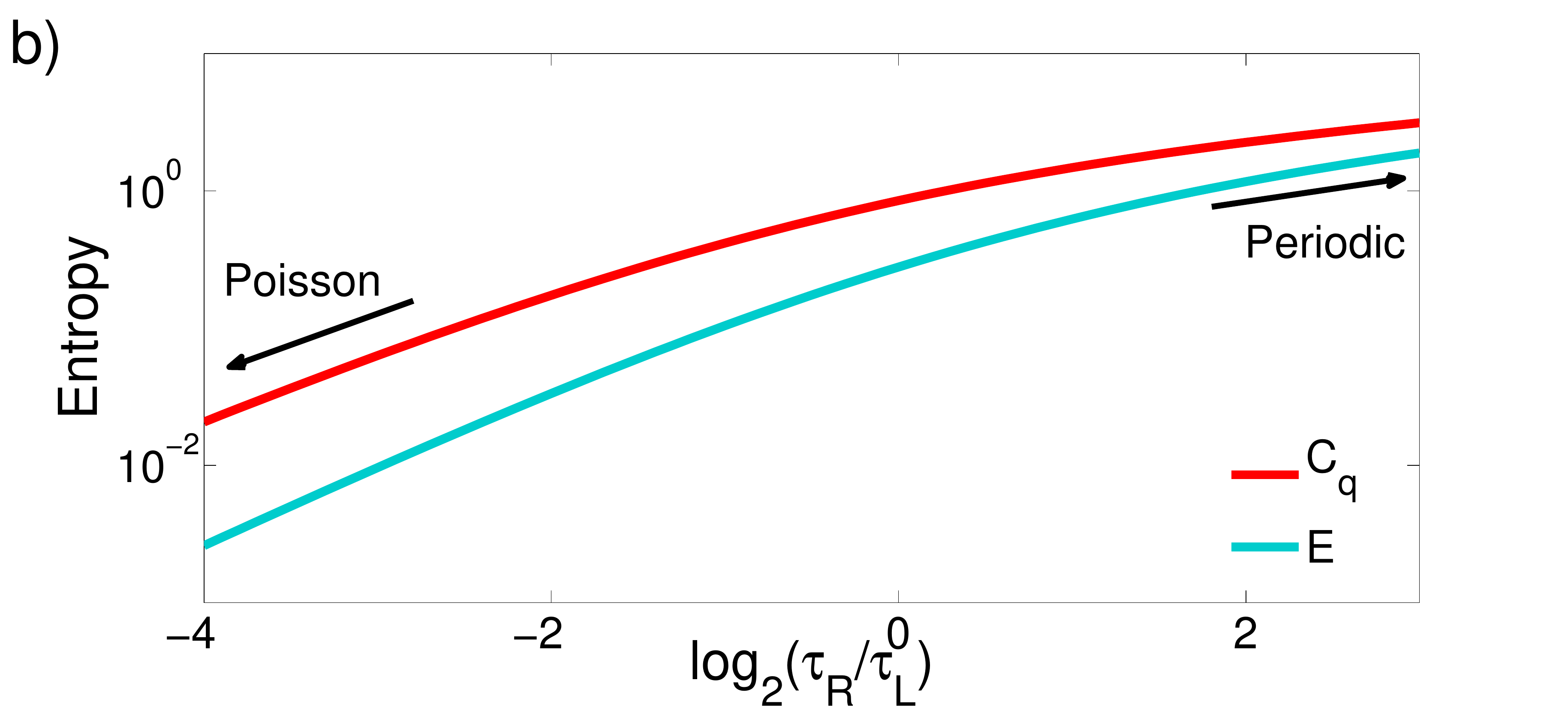}
\caption{{\bf Delayed Poisson process.} (a) The corresponding emission probability density for a delayed Poisson process with rest period $\tau_R$ and lifetime $\tau_L$. (b) Varying the ratio $\tau_R/\tau_L$ sweeps between a (memoryless) Poisson process and a periodic emission process, with a corresponding increase in the quantum memory $C_q$ required and excess entropy $E$ as the distribution becomes sharper. Here, $C_q$ is calculated approximately from a very fine $(2^{14}+1$ states) discretisation.}
\label{figdelay}
\end{figure}

For our second example, we consider a delayed Poisson process ($\phi(t)=(1/\tau_L)\exp(-(t-\tau_R)/\tau_L)$ for $t>\tau_R$ and 0 elsewhere), representing a process that exhibits an exponential decay with lifetime $\tau_L$, and a rest period $\tau_R$ between emissions [Fig.~5(a)], forming, for example, a very crude model of a neuron firing. For this emission distribution we find that $\mu=(\tau_L+\tau_R)^{-1}$, and $\Phi(t)=1$ for $t\leq\tau_R$ and $\exp(-(t-\tau_R)/\tau_L)$ for $t>\tau_R$. We can then show somewhat indirectly (see Appendix F) that the corresponding quantum memory requirement is bounded for finite $\tau_R/\tau_L$ (and vanishes as this ratio tends to zero), while in contrast $C_\mu$ is infinite whenever this is non-zero. Further, due to the timescale invariance of the quantum memory, $C_q$ depends only on this ratio, and not the individual values of $\tau_R$ and $\tau_L$. Varying this ratio allows us to sweep between a simple Poisson process with lifetime $\tau_L$, and a periodic process where the system is guaranteed to emit within an arbitrarily small interval at time $\tau_R$ after the last emission. The quantum memory $C_q$ correspondingly increases with this ratio as we interpolate between the two limits [Fig.~5(b)], with the pure Poisson process being memoryless, and a periodic process requiring increasing memory with the sharpness of the peak. We also plot the excess entropy, given by $E=\log_2(\tau_R/\tau_L+1)-\log_2e/(\tau_L/\tau_R+1)$, which exhibits similar qualitative behaviour.


{\bf Discussion.} We have shown that quantum devices can simulate models of continuous-time renewal processes with lower internal memory requirement than their corresponding optimal classical counterparts. Our examples evidence that this advantage can be arbitrarily large, compressing the need for an infinite classical memory into a finite quantum memory. Further, while we currently lack a proof, we suspect that this unbounded compression is a typical property of quantum machines. Our argument for this is as follows: Consider a discretised analogue of the process, with timesteps $\delta t$. When refining the discretisation to timesteps $\delta t/2$, we introduce an additional causal state for each one already existing. Classically, all these states are orthogonal, and the classical memory requirement increases, leading to the logarithmic divergence of $C_\mu$ as the timestep vanishes in the continuous limit. Contrarily, the overlap of adjacent quantum causal states $\ket{S(t)}$ and $\ket{S(t+\delta t)}$ is typically very large for small $\delta t$, and hence the additional interpolated states are very similar to the existing states. This overlap tends to one as the timestep vanishes, and thus as we tend to the continuum limit, the additional quantum causal states are essentially identical to those already considered, and hence the entropy increase with refinement should vanish. Thus we expect that, with the exception of pathological cases (such as when the distribution $\phi(t)$ involves arbitrarily sharp peaks), the quantum memory requirement will be finite.

It is prudent of course, to remark on the experimental feasibility of our proposal. Recent works have succeeded in realising quantum machines for discrete-time processes, and demonstrating their advantage over classical devices \cite{palsson2017experimentally}. Much effort has been spent on developing state-engineering protocols \cite{verstraete2009quantum, yi2012driven, hauke2013quantum, pedersen2014many, elliott2015multipartite}. Ultracold atoms in optical lattices \cite{lewenstein2012ultracold} may provide a route to realise discrete-time simulation,  and of particular promise with regards to our proposal for simulating continuous-time processes, photon pulses have been shaped over distances of several hundred metres \cite{nisbet2011highly}. This would allow for the digital simulation we propose, with the requirement of an additional control element needed to achieve the `real-time' analogue simulation. The on-going development of quantum technologies in a panoply of different systems holds much promise for the future implementation of our work.

Future theoretical work in this area can progress in many different directions, including the characterisation of other information-theoretic quantities \cite{james2011anatomy, marzen2014information} for the continuous-time quantum machines, the application to study real-world stochastic systems, and the extension of the protocol to design quantum simulators for models of more general continuous-time processes. Further, while a quantum advantage over classical simulators has been demonstrated, the general optimal construction of quantum machines is unknown, and a subject for future investigation.

{\bf Acknowledgements.} We thank Andrew Garner, Felix Binder and Suen Whei Yeap for useful discussions. This work was funded by Singapore National Research Foundation Fellowship NRF-NRFF2016-02. M.G.~is also financially supported by the John Templeton Foundation Grant 53914 ``Occam's quantum mechanical razor: Can quantum theory admit the simplest understanding of reality?" and the Foundational Questions Institute. T.J.E.~thanks the Centre for Quantum Technologies, National University of Singapore and the Department of Physics, University of Oxford for their hospitality.

\section*{Technical Appendix}

\appendix


{\bf A: Causal architecture for renewal processes.} For the purpose of determining their causal architecture, renewal processes can be classified as one of four types \cite{marzen2017informational}. Here we review this classification, and describe the corresponding structure of their causal states.

The first class of processes are termed \emph{eventually $\Delta$-Poisson}. These are processes which exhibit a periodic structure under a Poissonian envelope at long times. Specifically, an eventually $\Delta$-Poisson process is described by an emission probability of the form $\phi(t)=\phi(\tau+(t-\tau)\mathrm{mod}\Delta)\exp(-(t-\tau)/\tau_L)$ for times $t\geq\tau$, for some $\tau,\tau_L,\Delta\geq0$.

The second class of process we consider, \emph{eventually Poisson}, are a special case of the above, for which $\Delta=0$. In these processes, there is no structure at long times $t>\tau$ beyond the Poissonian decay. That is, specifically, for $t>\tau$ we have $\phi(t)=\phi(\tau)\exp(-(t-\tau)/\tau_L)$.

A yet further constrained form defines our third class, the familiar \emph{Poisson} process,  with $\Delta=0$ and $\tau=0$. For these processes, $\phi(t)=\exp(t/\tau_L)/\tau_L$. We find that the conditional emission probability density of this class is time-independent.

Finally, the fourth class, \emph{not eventually $\Delta$-Poisson}, encompasses all other processes not of the above forms.

We now present the causal architecture for each of the above process classes \cite{marzen2017informational}. We present this architecture in a different order to that in which we presented the processes, in order to introduce their different features in order of increasing intricacy. 

Recall that two times since last emission $t_{0^+}$ and $t_{0^+}'$ belong to the same causal state iff they have identical conditional probability distributions $P(T_{0^-}|T_{0^+})$ for the time to the next emission $t_{0^-}$. As a result of this, and because for a Poisson process the conditional emission probability density is time independent, all $t_{0^+}$ belong to a single causal state [Fig.~6(a)] for such processes.

\begin{figure}
\includegraphics[width=0.95\linewidth]{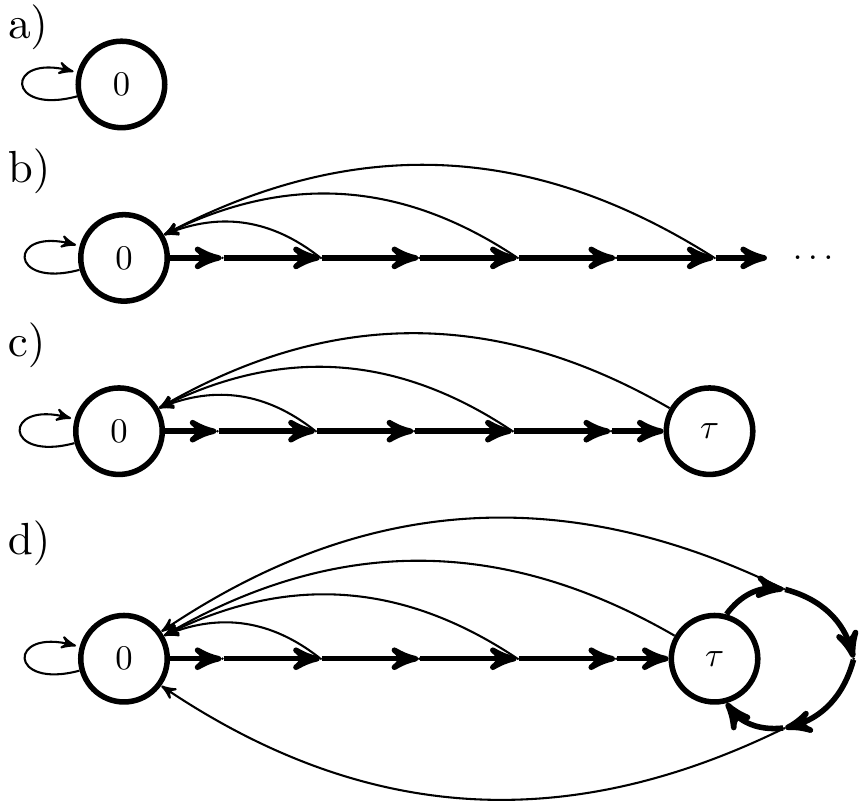}
\caption{{\bf Causal architecture of renewal processes.} The causal architecture depends on the class of the renewal process, with (a) a single causal state for Poisson processes, (b) an infinite continuum of causal states for not eventually $\Delta$-Poisson processes, and a hybrid of the two for (c) eventually Poisson and (d) eventually $\Delta$-Poisson processes.}
\label{figcausalarch}
\end{figure}

In stark contrast to this simplicity found for Poisson processes, it has been shown that for not eventually $\Delta$-Poisson processes no two different $t_{0^+}$ belong to the same causal state \cite{marzen2017informational}. Instead we have an infinite continuum of causal states, which the system traverses along between emissions, with all such continuum states returning to the same initial state immediately after an emission event (see Fig.~6(b)).

The remaining two classes of process are hybrids of the above structures. The eventually Poisson process begins with a continuum of distinct states for times $t_{0^+}<\tau$, after which it terminates in a single state upon reaching the Poissonian stage [Fig.~6(c)], as the conditional emission probability density becomes time-independent for $t_{0^+}\geq\tau$.

Finally, the eventually $\Delta$-Poisson process also begins with a continuum of distinct states for $t_{0^+}<\tau$, after which the system enters a periodic loop of continuum states of length $\Delta$, mirroring the periodicity of the conditional emission probability density for such processes at times $t_{0^+}\geq\tau$ (see Fig.~6(d)). At these long times, we do not need to track exactly how long it has been since the last emission, but merely how far into the current period we are.


{\bf B: Statistical complexity of renewal processes.} We now derive expressions for the statistical complexity of the different classes of renewal process. We do so by considering a discretised analogue \cite{marzen2015informational} of the continuum causal states, and take the limit of infinitesimal time intervals. 

Starting with the case of a Poisson process, due to the single causal state we need not track any information about the time since the last emission; we thus have $C_\mu=0$ and hence the process is memoryless.

We next jump to the case of a not eventually $\Delta$-Poisson process, where we have the perpetual continuum of states. Considering the discrete analogues $\sigma_t$ of the causal states $S_t$, we have states at each $t=n\delta t$ for $n\in\mathbb{N}$, where $\delta t$ is our discretised time interval. Consider now when the system is in causal state $\sigma_{n\delta t}$. In the next time interval $\delta t$ the system will either make an emission, or progress along the continuum to $\sigma_{(n+1)\delta t}$. This latter event occurs with probability $\Phi((n+1)\delta t)/\Phi(n\delta t)$ (that is, the conditional probability that the system does not emit before $(n+1)\delta t$ given that it did not emit before $n\delta t$), and we can consider the system to fictitiously emit a $\emptyset$ (null) symbol, representing the lack of a real emission. The alternative outcome is that the system actually emits a real symbol and returns to $\sigma_0$, which occurs with probability $1-\Phi((n+1)\delta t)/\Phi(n\delta t)$. This is illustrated in Fig.~7. Note that for a Poisson process, $\sigma_0$ is the lone causal state, and the discretised analogue consists of this single state with both transitions leading back to the same state, with the probabilities as given in the general case.

\begin{figure*}
\includegraphics[width=0.9\linewidth]{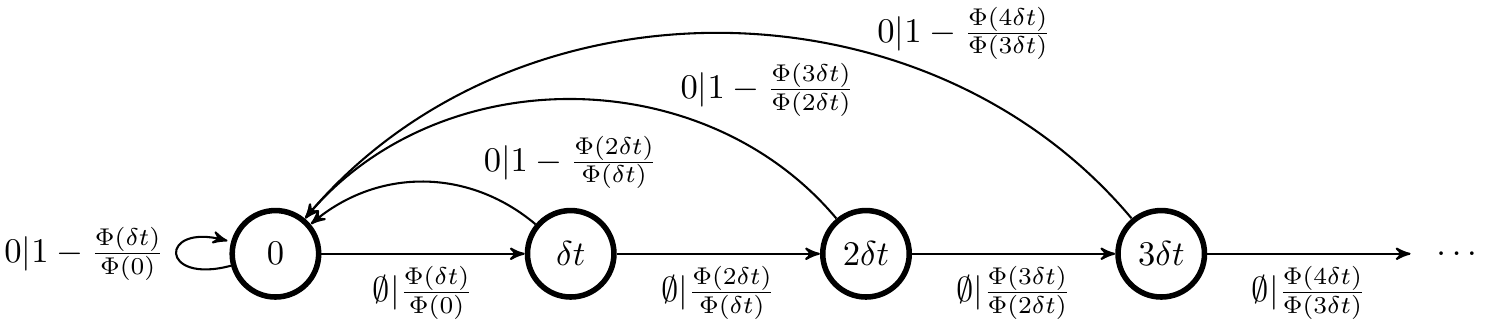}
\caption{{\bf Discretised analogue of a renewal process.} We can construct a discretised analogue of a renewal process, where at each time step $\delta t$ the system can either emit a symbol 0 and return to the initial causal state, or emit nothing (signified by a null symbol $\emptyset$) and proceed to the next state in the chain. The continuous-time scenario follows as the $\delta t\to0$ limit.}
\label{figdiscrete}
\end{figure*}

We can construct a transition matrix $T$ that describes the evolution of the system for each timestep. This matrix evolves a state $\omega$ according to $\omega(t+\delta t)=T\omega(t)$. The transition matrix for a not eventually $\Delta$-Poisson renewal process is given by
\begin{equation}
T=\begin{pmatrix}1-\frac{\Phi(\delta t)}{\Phi(0)}&1-\frac{\Phi(2\delta t)}{\Phi(\delta t)}&1-\frac{\Phi(3\delta t)}{\Phi(2\delta t)}&\ldots\\\frac{\Phi(\delta t)}{\Phi(0)}&0&0&\ldots\\ 0&\frac{\Phi(2\delta t)}{\Phi(\delta t)}&0&\ldots\\0&0&\frac{\Phi(3\delta t)}{\Phi(2\delta t)}&\ldots\\\vdots&\vdots&\vdots&\ddots\end{pmatrix}.
\end{equation}
The steady state $\pi$ is defined according to $\pi=T\pi$, and thus for $n\geq1$ we have $\pi(\sigma_{n\delta t})=(\Phi(n\delta t)/\Phi((n-1)\delta t))\pi(\sigma_{(n-1)\delta t}),$ and hence iteratively we find $\pi(\sigma_{n\delta t})=\Phi(n\delta t)\pi(\sigma_0)$. By integrating over $\Phi(t)$ to find the appropriate normalisation, we have that as $\delta t\to0$, $\pi(\sigma_0)\to\mu\delta t$, and hence $\pi(\sigma_t)=\mu\Phi(t)\delta t$. Inserting this into the Shannon entropy Eq.~\eqref{eqcmucont}, we have that the statistical complexity is given by
\begin{equation}
C_\mu=\lim_{\delta t\to0}-\sum_{n=0}^\infty\mu\Phi(n\delta t)\delta t\log_2(\mu\Phi(n\delta t)\delta t),
\end{equation}
which clearly diverges logarithmically as the argument of the logarithm vanishes (the number of terms in the sum grows linearly, while the prefactor to the logarithm decays linearly, thus effectively negating each other), and hence whenever such a continuum of causal states occurs (i.e.~any renewal process that is not Poisson), the classical memory requirement is infinite. As noted in the main text, the self-assembly of the quantum causal states means that we can effectively treat any renewal process as not eventually $\Delta$-Poisson in the quantum regime, and hence this steady state distribution is sufficient for our purposes. However, we will provide expressions for the steady state distributions and complexity of the eventually Poisson and eventually $\Delta$-Poisson processes for completeness.

For eventually Poisson processes, the continuum has a finite length, after which the system resides in the final causal state until emission. The probability density for the continuum states is as for the above case, and the probability of occupation of the final state can be determined by considering the average time spent in this state, given by the lifetime of the state $\tau_L$. Thus, the steady state occupation of this final state is $\pi(\sigma(\tau))=\mu\Phi(\tau)\tau_L$, and the corresponding statistical complexity of the process is 
\begin{align}
C_\mu=&\lim_{\delta t\to0}-\sum_{n=0}^{N-1}\mu\Phi(n\delta t)\delta t\log_2(\mu\Phi(n\delta t)\delta t)\nonumber\\&-\mu\Phi(\tau)\tau_L\log_2(\mu\Phi(\tau)\tau_L),
\end{align}
where $N=\tau/\delta t$.

Finally, for eventually $\Delta$-Poisson processes, we have two segments of continuum per emission; the initial line where each state is occupied at most once, and the periodic continuum where each state can be traversed multiple times per emission. There is a probability $\Phi(\tau)$ that the system reaches this periodic part on a given emission, and a probability $\Phi(\tau)\exp(-m\Delta/\tau_L)$ that it makes it through $m$ circuits of this periodic component. Thus, the occupation probability in the steady state of the periodic continuum state $\sigma_t$ ($\tau\leq t<\tau+\Delta)$ is $\pi(\sigma_t)=\Phi(t)\sum_{m=0}^\infty \exp(-m\Delta/\tau_L)=\Phi(t)/(1-\exp(-\Delta/\tau_L)$. This gives a statistical complexity of 
\begin{align}
C_\mu=\lim_{\delta t\to0}&-\sum_{n=0}^{N_1-1}\mu\Phi(n\delta t)\delta t\log_2(\mu\Phi(n\delta t)\delta t)\nonumber\\&-\sum_{n=0}^{N_2-1}\frac{\mu\Phi(\tau+n\delta t)\delta t}{1-e^{-\Delta/\tau_L}}\log_2\left(\frac{\mu\Phi(\tau+n\delta t)\delta t}{1-e^{-\Delta/\tau_L}}\right),
\end{align}
where $N_1=\tau/\delta t$ and $N_2=\Delta/\delta t$.

{\bf C: Memory of Decohered Model.} In the main text we claim that decoherence of the quantum causal states destroys the quantum advantage, and results in a memory cost $C_\mu$. We now prove this here by showing the probability distribution of the decohered states is identical to that of the steady states of the classical model, and hence has the same entropy. Consider the probability of the decohered state $S_D$ being $t$:
\begin{align*}
P(S_D=t)&=\int_0^\infty dt'P(S_D=t|T_{0^+}=t')P(T_{0^+}=t')\nonumber\\
&=\int_0^\infty dt'\frac{\phi(t+t')}{\Phi(t')}\mu\Phi(t')\nonumber\\
&=\mu\int_0^\infty dt'\phi(t+t')\nonumber\\
&=\mu\int_t^\infty dt'\phi(t')\nonumber\\
&=\mu\Phi(t)=P(T_{0^+}=t).
\end{align*}
Note that the resultant model using these decohered states, while not providing any memory savings, does contain oracular information and so would not be a predictive model.


{\bf D: Timescale Invariance of Quantum Memory.} Here we show that the memory requirement of q-machines is invariant under rescaling of the time variable in the emission probability density. Consider such a rescaling $t\to\alpha t=z$. The wavefunction scales $\psi(t)\to\sqrt{\alpha}\psi(z)$ (the factor in front due to normalisation), and the mean firing rate changes $\mu\to\alpha\mu$. Putting this into the charateristic equation Eq.~\eqref{eqcharacteristic}, and using the substitution $dt=dz/\alpha$, we have
\begin{align}
&\mu\int_0^\infty\!\!\!\! db\int_0^\infty\!\!\!\! dt\psi(t+a)\psi(t+b)f_n(b)=\lambda_nf_n(a) \nonumber\\
\to&\alpha\mu\int_0^\infty\!\!\!\! db\int_0^\infty\!\!\!\! dt \alpha\psi(z+z_a)\psi(z+z_b)f_n(b)=\lambda_nf_n(a)\nonumber\\
=&\mu\int_0^\infty\!\!\!\! dz_b\int_0^\infty\!\!\!\! dz \psi(z+z_a)\psi(z+z_b)f^{[\alpha]}_n(z_b)=\lambda_nf_n^{[\alpha]}(z_a),
\end{align}
where we have defined the function $f_n^{[\alpha]}(t)=f_n(t/\alpha)$. Thus, after the rescaling, the characteristic equation is of the same form, solved by the same eigenvalues $\lambda_n$ and the rescaled (and renormalised) eigenfunctions $f_n^{[\alpha]}$. As the memory stored depends only on the $\lambda_n$, it is hence unchanged: $C_q$ is timescale invariant.


{\bf E: Technical details for uniform emission probability example.} Here we provide details of the derivation of the boundedness of the quantum memory for the uniform emission probability case, along with the other associated quantities. Starting with the definition of the process
\begin{equation}
\phi(t)=\left\{\begin{array}{ll} \frac{1}{\tau}  & 0\leq t<\tau \\ 0 & t \geq \tau\end{array}\right. ,
\end{equation}
we can straightforwardly obtain that $\mu^{-1}=\int_0^\tau (t/\tau)dt=\tau/2$, and 
\begin{equation}
\Phi(t)=\int_t^\infty\phi(t')dt'=\left\{\begin{array}{ll} 1-\frac{t}{\tau}  & 0\leq t<\tau \\ 0 & t \geq \tau\end{array}\right. .
\end{equation}
As the time between emissions is guaranteed to be less than $\tau$, we can consider quantum causal states only within the interval $[0,\tau)$.

The form of the quantum causal states $\ket{S_t}=\int_0^{\tau-t}dt'(1/\sqrt{\tau-t})\ket{t'}$ follows directly from the definition Eq.~\eqref{eqquantumcausal}, and we can construct the appropriate characteristic equation for the process:
\begin{equation}
\frac{2}{\tau}\int_0^\tau db\left(1-\frac{\mathrm{max}(a,b)}{\tau}\right)f_n(b)=\lambda_n f_n(a).
\end{equation}
Taking the second derivative of both sides of this equation, we find that
\begin{equation}
\lambda_n \frac{d^2 f_n}{d t^2}=-\frac{2}{\tau^2}f_n,
\end{equation}
and hence the eigenfunctions are of the form $f_n(t)=A\exp(ik_nt)+B\exp(-ik_nt)$, with $k_n^2=2/(\lambda_n\tau^2)$. We must now determine the values of $k_n$ that are valid, by substituting this solution into the original integral equation. Doing so, we find that for consistency, the following conditions must be satisfied: $A=B;$ and $\cos(k_n\tau)=0$. These are satisfied by $k_n\tau=(n-1/2)\pi$ for $n\in\mathbb{Z}^+$ (zero and negative integer values of $n$ produce solutions with the same eigenfunctions, by symmetry). Thus, we have that
\begin{equation}
\lambda_n=\frac{2}{\left(n-\frac{1}{2}\right)^2\pi^2}\quad n\in\mathbb{Z}^+.
\end{equation}
We now wish to find the Shannon entropy of the $\lambda_n$. We first show that this entropy is bounded by using an integral test \cite{knopp1956infinite} for convergence, and then use this to provide a bounded range for $C_q$. Define the function $\zeta(n)=-\lambda(n)\log_2\lambda(n)$, where $\lambda(n)=2/(n-1/2)^2\pi^2$, the interpolated continuous analogue of the eigenvalues $\lambda_n$. The sum of $\zeta(n)$ over positive integers $n$ gives the quantum memory cost, and we note that all such values of $\zeta(n)$ are finite. Further, we note that with the exception of $n=1$, they satisfy $\zeta(n)>\zeta(n+1)$ at these integer values, and hence the function is monotonically decreasing for $n>2$ (specifically, the continuous function is decreasing for $n>1/2+\exp(\ln2/2+2/\pi)/\pi$). Since the integral test requires the terms to be monotonically decreasing, we can sum up the terms to some finite $N>1$, and then show the remainder of the terms converge.

Define the integral $I(N)=\int_N^\infty \zeta(x)dx$. We find that
\begin{equation}
I(N)=\frac{2\left(\log_2\left(\left(N-\frac{1}{2}\right)^2\pi^2\right)-1\right)}{\pi^2\left(N-\frac{1}{2}\right)}+\frac{4}{\pi\left(N-\frac{1}{2}\right)\ln2},
\end{equation}
which is finite for all integer $N>1$, and hence the sum converges, implying a finite value for $C_q$. We can also use this to bound the value of the sum from $N$ to $\infty$ as being between $I(N)$ and $I(N)+\zeta(N)$. For $N=2$, this allows us to bound $1.1046\lesssim C_q
\lesssim 1.4174$. By calculating additional terms in the sum prior to taking the bound, we can tighten this further; for $N=10^6$, we find that $C_q\approx1.2809$, with the additional neglected terms contributing $\mathcal{O}(10^{-5})$ as a correction.

As the excess entropy is a property of the process, rather than the simulator, we can use the same formula as for classical models \cite{marzen2017informational}. This reads
\begin{equation}
\label{eqexcessentropy}
E=\mu\int_0^\infty \!\!\!\!t\phi(t)\log_2(\mu\phi(t))dt-2\mu\int_0^\infty\!\!\!\!\Phi(t)\log_2(\mu\Phi(t))dt.
\end{equation}
Putting the appropriate expressions into this equation, we find that for the uniform emission probability renewal process that the excess entropy is given by $E=\log_2e-1\approx0.4427$, which as expected is less than $C_q$.


{\bf F: Technical details for delayed Poisson process example.} Here we provide the corresponding details of the derivations for the delayed Poisson process example. Recall that the process is defined by the emission probability density
\begin{equation}
\phi(t)=\left\{\begin{array}{ll} 0  & 0\leq t\leq\tau_R \\ \frac{1}{\tau_L}e^{-(t-\tau_R)/\tau_L} & t > \tau_R\end{array}\right. ,
\end{equation}
from which it is straightforward to calculate $\mu^{-1}=\tau_L+\tau_R$, and 
\begin{equation}
\Phi(t)=\left\{\begin{array}{ll} 1  & 0\leq t\leq\tau_R \\ e^{-(t-\tau_R)/\tau_L} & t > \tau_R\end{array}\right. .
\end{equation}

We can exploit the causal architecture discussed in Appendix A, and identify this as an eventually Poisson process. We thus have a continuum of causal states for $0\leq t_{0^+}<\tau_R$, and a single causal state for $t_{0^+}\geq\tau_R$. From the mean firing rate and survival probability, we can see that the probability density for all continuum causal states in the steady state is given by $(\tau_L+\tau_R)^{-1}$, while the probability that the eventually Poisson causal state is occupied is $\tau_L/(\tau_L+\tau_R)$.

\begin{figure}
\includegraphics[width=\linewidth]{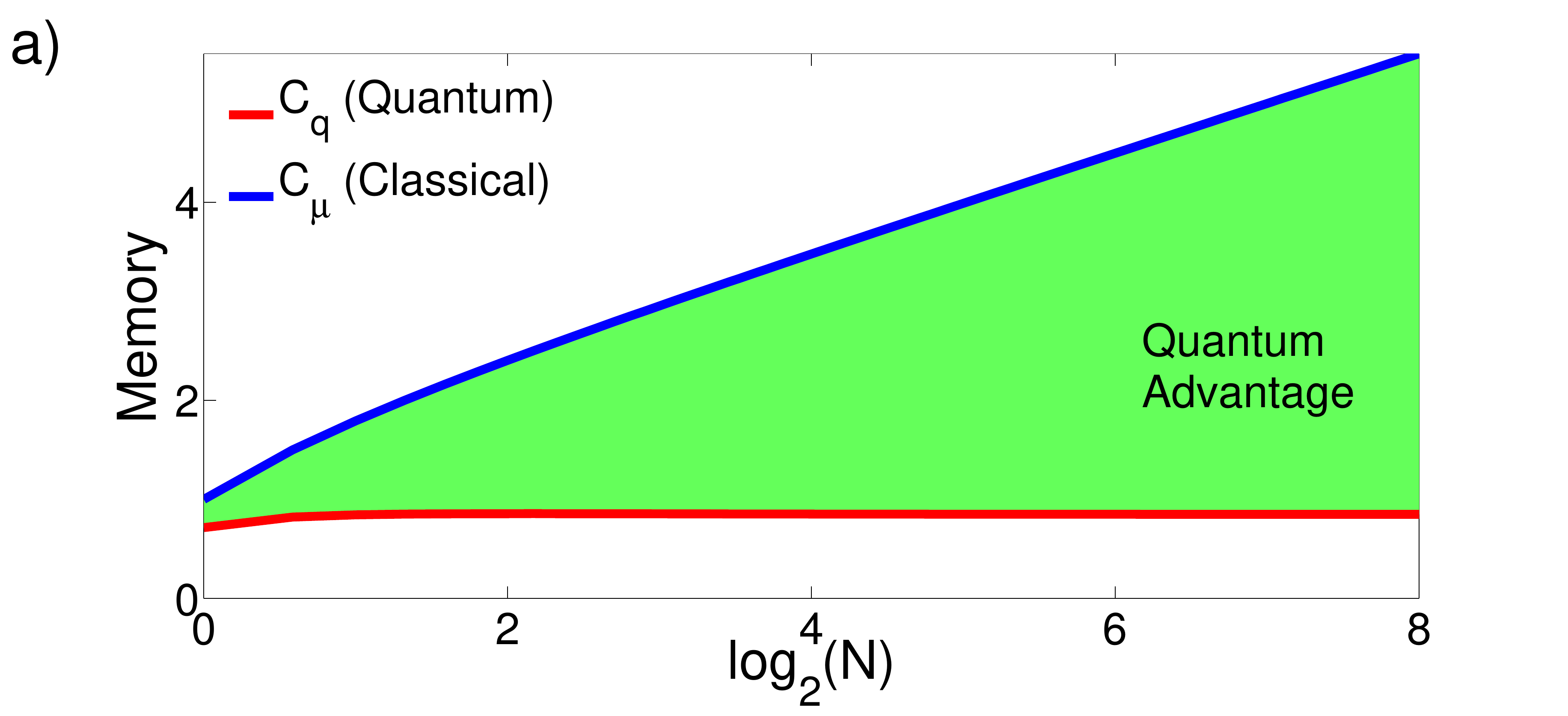}
\includegraphics[width=\linewidth]{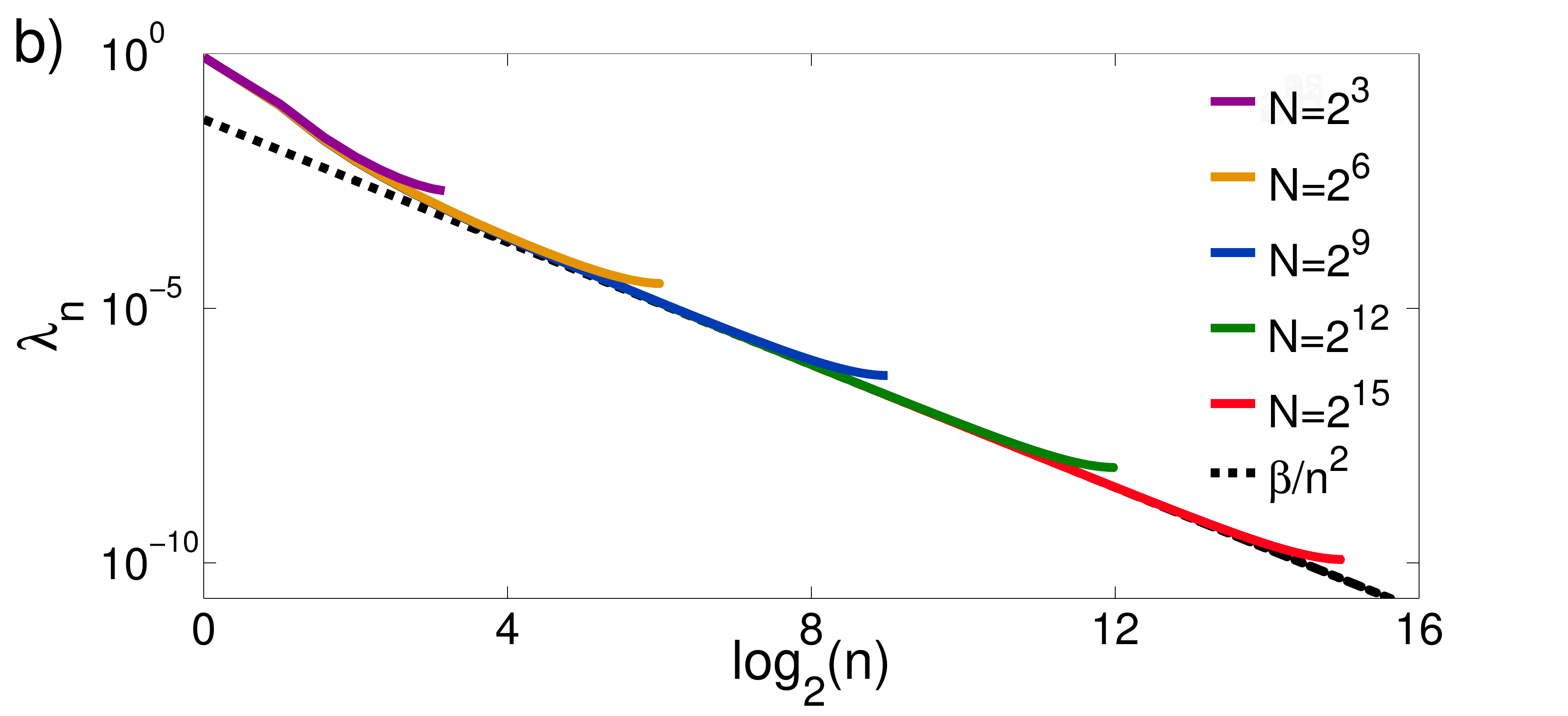}
\caption{{\bf Convergence of quantum memory for delayed Poisson process.} (a) The classical memory requirement $C_\mu$ for the delayed Poisson process diverges logarithmically with finer discretisation ($N+1$ states), while the quantum memory $C_q$ appears to converge to a finite value. (b) Inspection of the eigenvalues of increasingly finer discretisation of the q-machine for the delayed Poisson process shows that the eigenvalues appear to fall off with a $1/n^2$ dependence. Plots shown for $\tau_R/\tau_L=1$, and $\beta$ is a normalisation constant chosen such that $\sum_{n=101}^\infty \beta/n^2=\sum_{n=101}^{N+1}\lambda_n$ for the $N=2^{15}$ case (eigenvalues ranked largest to smallest).}
\label{figdelayconverge}
\end{figure}

While we can approximately determine $C_q$ by considering discretised time intervals $\delta t\ll\tau_L,\tau_R$ and see that the quantum memory appears to converge to a finite value (see Fig.~8), it is not a simple task to find an analytical expression for the continuous time limit. Instead, we shall prove boundedness of the quantum memory by considering a less efficient encoding of the causal states, and proving that this suboptimal encoding scheme has a bounded $C_q$. Specifically, we encode the eventually Poisson state $\ket{S_{\tau_R}}$ to be orthogonal to the continuum states. The density matrix is now block-diagonal, with one block for the continuum states, and a single element for the eventually Poisson state, and thus the total entropy is the sum of the entropies of the two blocks. The eventually Poisson state block contributes a finite amount (as it is a single element), and we shall now show that the contribution from the continuum block is also finite.

We can use the characteristic equation Eq.~\eqref{eqcharacteristic} to find the entropy contribution from the continuum block. We find that the overlap of two quantum causal states is given by $\braket{S_a}{S_b}=\exp(-|a-b|/2\tau_L)$, and hence
\begin{equation}
\label{eqintegraldecay}
\frac{1}{\tau_L+\tau_R}\int_0^{\tau_R} db e^{-|a-b|/2\tau_L} f_n(b)=\lambda_n f_n(a).
\end{equation}
Differentiating twice, we obtain
\begin{equation}
\frac{d^2f_n}{dt^2}=-\frac{1}{4\tau_L}\left(\frac{4}{\lambda_n(\tau_L+\tau_R)}-\frac{1}{\tau_L}\right)f_n,
\end{equation}
and so as with the previous case, the eigenfunctions are of the form $f_n(t)=A\exp(ik_nt)+B\exp(-ik_nt)$, now with
\begin{equation}
k_n^2=\frac{1}{4\tau_L}\left(\frac{4}{\lambda_n(\tau_L+\tau_R)}-\frac{1}{\tau_L}\right).
\end{equation}
Again, we substitute into the original integral equation Eq.~\eqref{eqintegraldecay}, which results in the consistency equations $(1-2ik_n\tau_L)A=(1+2ik_n\tau_L)B$ and Im($(1+2ik_n\tau_L)^2\exp(ik_n\tau_R)$)=0. Thus, the valid $k_n$ satisfy 
\begin{equation}
\label{eqdelayedconsistency}
\tan(k_n\tau_R)=\frac{4\tau_L k_n}{4\tau_L^2 k_n^2 -1},
\end{equation}
which, with one exception has one solution in each interval $[m\pi,(m+1)\pi)$, $m\in\mathbb{N}$. The exception is during the interval in which $4\tau_L^2k_n^2=1$, in which case there may be two solutions. For values of $k_n$ for which $4\tau_L^2k_n^2\gg1$, this is approximately satisfied by $k_n\tau_R=n\pi$, for $n\in\mathbb{Z}^+$, which leads to corresponding eigenvalues
\begin{equation}
\lambda_n=\frac{4}{\left(\frac{\tau_R}{\tau_L}+1\right)\left(4n^2\pi^2\frac{\tau_L^2}{\tau_R^2}+1\right)}.
\end{equation}
 Strictly, these approximate eigenvalues are overestimations, as the solutions to Eq.~\eqref{eqdelayedconsistency} for large $k_n$ are slightly larger than $n\pi/\tau_R$. However, as these $\lambda_n\ll1$, this also overestimates their contribution to the entropy. We further note that when $4n^2\pi^2\tau_L^2/\tau_R^2\gg1$ (i.e. for sufficiently large $n$), the $\lambda_n$ scale approximately as $1/n^2$. Again, this simplification overestimates the eigenvalues, and their contribution to the entropy. We can then break up the entropy into two parts; that from the finite number of terms preceeding the values of $n$ for which the above approximations are valid (which, due to the finite number of terms, gives a finite contribution), and those that come from the terms in which we have such large $n$. These latter terms also have a finite contribution to the entropy due to their $1/n^2$ scaling, and hence the total entropy is finite. This completes our proof that the quantum memory requirement for the delayed Poisson process is finite, though unlike the previous example we do not have an analytical expression for this value.

We can again calculate the excess entropy using Eq.~\eqref{eqexcessentropy}, and after some straightforward (if somewhat tedious) integration we obtain that $E=\log_2(\tau_R/\tau_L+1)-\log_2e/(\tau_L/\tau_R+1)$, which lies below and follows similar behaviour to the memory requirement $C_q$.

\end{document}